\documentclass[fleqn,usenatbib]{mnras}

\usepackage{newtxtext,newtxmath}

\usepackage[T1]{fontenc}
\usepackage{ae,aecompl}
\usepackage{url}
\usepackage{hyperref}

\usepackage{graphicx}	
\graphicspath{ {images/} }
\usepackage{amsmath}	
\usepackage{amssymb}	

\title[Black Hole Detection with LISA]{Evaluating Black Hole Detectability with LISA}

\author[Michael L. Katz and Shane L. Larson]{
Michael L. Katz,$^{1,2}$\thanks{E-mail: mikekatz04@gmail.com}
Shane L. Larson,$^{1,2}$
\\
$^{1}$Department of Physics and Astronomy, Northwestern University,
    Evanston, IL 60208, United States\\
$^{2}$Center for Interdisciplinary Exploration and Research in Astrophysics (CIERA), Evanston, IL 60208, United States\\
}

\date{Accepted 2018 November 27. Received 2018 November 26; in original form 2018 July 6}

\pubyear{2018}

\begin{document}
\label{firstpage}
\pagerange{\pageref{firstpage}--\pageref{lastpage}}
\maketitle

\begin{abstract}

We conduct an analysis of the measurement abilities of distinctive LISA detector designs, examining the influence of LISA's low-frequency performance on the detection and characterization of massive black hole binaries. We are particularly interested in LISA's ability to measure massive black holes merging at frequencies near the low-frequency band edge, with masses in the range of $\sim 10^6-10^{10}M_\odot$. We examine the signal-to-noise ratio (SNR) using phenomenological waveforms for inspiral, merger, and ringdown over a wide range of massive black hole binary parameters. We employ a broad palette of possible LISA configurations with different sensitivities at low frequencies. For this analysis, we created a tool\footnotemark that evaluates the change in SNR between two parameterized situations. The shifts in SNR are computed as gains or losses as a function of binary parameters, and graphically displayed across a two dimensional grid of parameter values. We illustrate the use of this technique for both parameterized LISA mission designs, as well as for considering the influence of astrophysical parameters on gravitational wave signal models. In terms of low-frequency sensitivity,  acceleration noise or armlength is found to be the most important factor in observing the largest massive black hole binaries, followed by break frequency and then spectral index. LISA's ability to probe the astrophysical population of $\sim10^7-10^9M_\odot$ black holes is greatly influenced by these aspects of its sensitivity. The importance of the constituent black hole spins is also highlighted.   
\end{abstract}

\begin{keywords}
gravitational waves -- quasars: supermassive black holes
\end{keywords}



\footnotetext{BOWIE, \url{https://github.com/mikekatz04/BOWIE}}
\section{Introduction}

The success of LISA Pathfinder and the selection of the Laser Interferometer Space Antenna (LISA) for the European Space Agency's L3 mission has reinvigorated interest in the science that a LISA mission will enable, particularly with regard to massive black hole binaries in the Universe. LISA is slated for launch in the early 2030s, and is currently in early design studies. The science ability of LISA will be a strong function of the overall shape of its sensitivity. As the mission design matures towards its final design, construction, and launch, it will be necessary to more fully understand what the source measurement limitations of the observatory are and how those limits influence the science returns of the mission. In this paper, we consider how choices for the low-frequency performance of LISA influence our ability to detect and characterize massive black hole binaries.

For this study, we consider intermediate mass black holes (IMBH) with masses on the order of $\sim10^2-10^5M_\odot$ and massive black holes (MBH) with masses on the order of $\sim10^6-10^{10}M_\odot$. MBHs exist in the centre of most galaxies of considerable size, having been inferred by electromagnetic observations of Galactic centres \citep{Kormendy1995}, including some pairs of MBHs forming binaries (\citeauthor{Graham2015} \citeyear{Graham2015}; \citeauthor{Bansal2017} \citeyear{Bansal2017}). Black holes of $\sim10^5M_\odot$ are beginning to be discovered in dwarf galaxies using similar electromagnetic techniques \citep{Reines2013}. Although a few groups have proposed electromagnetic detections of IMBHs (e.g. \citeauthor{Lin2018} \citeyear{Lin2018}), their existence remains uncertain without clear observational evidence. The Galactic centre black hole with the best inferred mass observations to date has been the Milky Way central MBH, constrained to a mass of $\sim4\times10^6M_\odot$ \citep{BoehleGhez2016}. MBHs similar in size to the Milky Way central black hole are prime targets for the LISA mission as these MBHs, when existing in binaries, merge at frequencies where LISA is most sensitive. These MBH binary coalescences result from mergers of their host galaxy haloes. Over time, the two MBHs sink to the centre of the primary halo via different dynamical processes  (\citeauthor{Haiman2009} \citeyear{Haiman2009}, \citeauthor{Dosopoulou2017} \citeyear{Dosopoulou2017}, \citeauthor{PTA-illustris} \citeyear{PTA-illustris}, \citeauthor{Rasskazov2} \citeyear{Rasskazov2}). 

Currently, knowledge of MBHs is limited to those that are close enough for dynamical measurements with surrounding matter (see \citeauthor{Kormendy2013} \citeyear{Kormendy2013} for review) or occur as active Galactic nuclei (AGN), which are visible out to higher redshifts \citep{Veron-Cetty2010}. \citet{Rasskazov1} suggest previous electromagnetic measurements of the galaxy bulge mass to MBH mass relation was overestimated by factors of 2 - 3, which has direct implications for Pulsar Timing Array (PTA) predictions and the MBH mass spectrum in general. Gravitational wave observations of merging MBH binaries will provide an independent way of probing the mass spectrum in environments containing these systems. If they are luminous enough, astrophysical interactions in MBH binaries could be more fully characterized using combined GW and EM observations, particularly with regard to accretion processes. Additionally, LISA will measure the luminosity distance of binary MBHs. Combining this information with the electromagnetic measurement of a binary's redshift can give the Hubble parameter \citep{Schutz1986}. LISA measurements alone can also constrain other cosmological parameters via statistical methods \citep{Petiteau2011}. Similarly, MBH spins can provide strong probes of general relativity and cosmological theory, while possibly shedding light on the interactions of particle theory with general relativity near a black hole's event horizon (\citeauthor{Gair2013} \citeyear{Gair2013}, \citeauthor{Barausse2015} \citeyear{Barausse2015}).

\subsection{Low-Frequency Band Edge}

In this paper we will examine the low-frequency band edge shape of the LISA sensitivity curve and its effects on gravitational wave measurements of MBH binaries. \emph{Analyzing the band edge helps us understand LISA's ability to measure populations of binaries evolving or merging at frequencies just below the LISA sensitivity band, showing that the relevant MBH mass ranges of interest in LISA observations can be as large as $10^9M_\odot$}. We perform numerical analysis for various sensitivity curves on a full range of MBH parameters independent of any particular model of MBH binary populations or evolution. Therefore, we strive to answer the following question: what is the ability of various LISA configurations to detect any model-independent merging binary with reasonable parameters? Previous work has considered binaries merging over the centre of the LISA band, where it is most sensitive, and rate predictions for different sensitivities based on population models, without specific differences in the band edge shape \citep{Klein2016}. Similarly, \citet{Berti2016} analysed ringdown signals over various sensitivities curves analytically and without specific band edge changes. 

Measurement of IMBHs are mainly affected by the high frequency performance of the sensitivity curve. IMBH signals are weaker at frequencies where they cross over into the LISA sensitivity band, likely remaining unresolved. While the low-frequency band edge shape can have some influence on IMBH analysis, the focus in this work will be primarily on MBH systems. Estimates for the detection rates of MBH mergers range from 1 to 100 per year. For further discussion on related detection rate predictions based on specific MBH formation models, see \citet{Sesana2011}, \citet{Berti2016}, \citet{Klein2016}, and \citet{Salcido2016}.

The goal of this paper is a framework to efficiently consider how LISA sensitivity performance changes the regions of parameter space that can be effectively probed by LISA observations.

\subsection{Coalescence Stages}

During a binary black hole coalescence, three sequential stages are observable in the gravitational wave data. These are inspiral, merger, and ringdown. Inspiral occurs when two black holes form a bound pair and begin to radiate away their orbital angular momentum and energy via gravitational radiation. This occurs at separations less than thousands or hundreds of Schwarzschild radii, at sub-parsec distances \citep{PTA-illustris}. As the black holes move closer together, the frequency of the gravitational radiation increases. This increase in frequency is referred to as ``chirping.'' The rate at which a binary chirps is given by
\begin{equation}
    \dot{f} = \frac{96}{5}\frac{c^3}{G}\frac{f}{M_c}\left(\frac{G}{c^3}\pi fM_c\right)^{8/3},
\end{equation}
where $M_c=(m_1m_2)^{3/5}/(m_1+m_2)^{1/5}$ is the chirp mass. The chirp is the rate at which the frequency evolves, and defines how long the source dwells in a given frequency bin. In the context of LISA observations, the time in band at specific frequencies influences our ability to characterize a source depending on the overall shape of the instrument sensitivity. Over much of the time a binary inspirals in the LISA band, the evolution is well approximated by a post-Newtonian inspiral.

Following the inspiral stage, merger occurs when the two black holes have reached a separation smaller than the Innermost Stable Circular Orbit (ISCO) \citep{FlanaganHughes1}. At this point, the post-Newtonian treatment breaks down and numerical relativity is needed to understand the behavior of the binary. This stage will provide a strong signal if it occurs in the LISA sensitivity band.

Once the two black holes have combined to form a single event horizon, the ringdown begins. The ringdown is caused by the remnant black hole radiating deviations from an axi-symmetric state until it becomes a Kerr black hole. This stage is addressed using black hole perturbation theory, and is well understood in the confines of General Relativity. This treatment involves different spherical harmonic modes of the black hole, where each mode of the ringing black hole will emanate gravitational waves at specific frequencies \citep{Berti2006}. The dominant mode is expected to be the $l=m=2$ spherical harmonic mode \citep{FlanaganHughes1}. However, subdominant modes of the ringdown can be measured if the signal conditions are right. This practice is known as ``black hole spectroscopy'' (\citeauthor{Berti2006} \citeyear{Berti2006}, \citeauthor{Berti2016} \citeyear{Berti2016}, \citeauthor{Baibhav2018} \citeyear{Baibhav2018}).

A changing sensitivity curve affects two main aspects related to the detection of an inspiral signal. For binaries that merge near the band edge, a change in detector performance can result in completely loosing or gaining a signal from the final stages of inspiral. The second aspect additionally involves binaries that merge at higher frequencies away from the band edge. The changing of the low-frequency band edge of the sensitivity curve can, in general, reduce or increase the number of inspiral cycles and accumulated signal-to-noise ratio (SNR, also referred to as ``$\rho$'') measured for a binary evolving in the LISA band. The  merger and ringdown signal can be detected even for a binary that reaches its merger frequency prior to entering the LISA band if its merger and/or ringdown spectrum reach across the low-frequency band edge. \emph{Adjusting the low-frequency properties of the noise curve will directly affect SNR measurements of merger and ringdown signals resulting in direct implications for the accumulation of MBH sources by LISA}.

\section{Methods}
\subsection{Graphically Comparing Detection Measures}

In previous work, sensitivity curve and binary configuration detection analysis have been illustrated with contour line plots or filled contour plots: one parameter on each axis and the SNR as the contour value. We will refer to filled SNR contour plots as ``waterfall'' plots. These plots allow the user to analyse the absolute SNR predicted for one specific sensitivity curve. This has been used previously to assess the ability of specific configurations to meet the requirements for the LISA mission (see figure 3 in \citeauthor{LISAMissionProposal} \citeyear{LISAMissionProposal}). Another type of plot commonly used is the ``horizon'' plot. Horizon plots can compare many configurations by showing a single contour line for each configuration. These are referred to as ``horizon'' plots because they are generally used to illustrate the limiting redshifts or luminosity distances of various sensitivity curves, therefore showing the entirety of the detectable parameter space for each configuration. We use horizon plots to show two types of graphical analyses. The first is the traditional type where limiting redshift contours are shown for an SNR indicating the minimal requirement for detectability. Usually, an approximate SNR cut between 5 and 10 is applied as the limiting value for sources to be resolvable to a degree to which parameterizing the source properties will be possible. In this paper, we use an SNR cut of $\rho=5$ to highlight sources near the threshold of detection. The second type of horizon plot we use shows contours for an SNR different from this limiting value. 

Visual comparison of horizon or waterfall plots succinctly captures performance or parameter sensitivity estimates over a broad range of plausible configurations or parameter values. We introduce a comparison between similar filled contour plots called a ``ratio plot,'' which takes the ratio of a detection statistic between two parameterized situations and plots colored contours based on this ratio. For our work, the SNR is used as the detection statistic. We chose to use the SNR because it is a good representation of detectability and, with simplified analytical estimates of Fisher matrix analysis, can translate to roughly determining parameter estimation capabilities. Direct computation of Fisher matrices is beyond the scope of this work. The ratio contours show where sensitivity decreases (blue) and increases (red) for a given comparison. Our ratio plots will show contours based on the value of $\rho_1/\rho_2$, where $\rho_i$ represents the SNR from one binary and sensitivity configuration. If either $\rho_1$ or $\rho_2$ is less than 5, the ratio value is not shown, leaving this area as white space. In other words, we only show these contours when a source is detectable by both configurations. This provides a clear delineation of behaviors exhibited in the comparison, showing exactly how the detection statistic improves or diminishes in different directions in the parameter space, and by what factor the improvement is growing or diminishing (represented by the gradient in the contours).

In addition to ratio contours, we also show areas referred to as ``loss/gain'' contours on the ratio plots. When comparing one sensitivity curve to another, we use ``gain'' and ``loss'' to indicate sources that would or would not appear in a LISA catalog, defined by an SNR threshold; however, in order to appear within the contour area, one configuration must achieve the SNR threshold while the other does not. To further understand these constructions, we can look at an example for possible configurations A and B with an SNR cut of 5. Here, A is the baseline; B is an alternative to be compared to A. For the first example, we look at a binary where $\rho_A<5$ and $\rho_B>5$. In this case, B measures the binary while A does not. Therefore, this binary will appear within a solid loss/gain contour as a ``gained'' source by B compared to A. When the opposite is true, $\rho_A>5$ and $\rho_B<5$, the binary will appear in a dashed contour as a ``lost'' source by B compared to A. If $\rho_A>5$ and $\rho_B>5$, or $\rho_A<5$ and $\rho_B<5$, the binary will not appear within a loss/gain contour. However, if $\rho_A>5$ and $\rho_B>5$, this binary will appear in a ratio contour because the source is detectable by both configurations. With the combination of ratio and loss/gain contours, we show how two configurations differ in which sources they can detect (loss/gain contours), as well as how their measurement strength compares for sources detectable by both configurations (ratio contours). See the figures below for examples of these constructions. 

\subsection{BOWIE}

With the tool, Binary Observability With Illustrative Exploration (BOWIE; \citeauthor{BOWIE} \citeyear{BOWIE}), this type of analysis can be extended to encompass many visual representations across desired parameters and/or sensitivity curves. By arranging ratio, waterfall, and horizon plots in a deliberate manner, a detailed analysis can be performed in many different ways.

BOWIE is a flexible tool to be used in many areas of gravitational wave analysis. BOWIE is made up of two main parts. The first uses phenomenological  waveforms from binary black hole coalescence to generate SNR grids; the second part takes these SNR grids and produces plots like those shown below. 

The grid generator is not specific to black holes related to LISA analysis, and can support binary black hole analysis in any part of the gravitational wave spectrum. BOWIE can be used for low-frequency MBH inspirals for Pulsar Timing Array searches, or stellar mass black hole mergers observable by LIGO. Similarly, BOWIE is directly applicable to analysis for future gravitational wave observatories. 

The plotting side of BOWIE has many extensions. We show below BOWIE can create sensitivity comparisons with a given set of binary parameters. Additionally, it can make comparisons over binary parameters, fixing the sensitivity. The user can also combine sensitivity and binary analysis because the structure of the package allows for direct choice of each plot shown within an adjustable overall configuration.

Separating the plotting module allows for extension beyond binary black hole analysis, therefore, opening this tool to any noise infused signal. BOWIE takes user-supplied SNR (or any other detection metric) grids as input, producing customizable outputs for a variety of applications.

BOWIE is written in the Python programming language and provides an easy to use interface in Python. It contains three packages: a \textit{PhenomD} (\citeauthor{Khan2016} \citeyear{Khan2016}; \citeauthor{Husa2016} \citeyear{Husa2016}) waveform generator and SNR calculator referred to as \textit{pyphenomd}; an SNR grid generation package called \textit{bowie\_gencondata}; and the package used to create the plots shown in this paper, \textit{bowie\_makeplot}. For both the SNR grid generation and plotting packages within BOWIE, the user supplies a Python Dictionary object to a Python function, detailing preferences for output. The Dictionary setup is designed to be user friendly for users who have minimal Python experience. The plotting module also allows for familiar Python users to edit their plot output within the \textit{pyplot} interface from the Matplotlib library. The \textit{pyphenomd} package provides a \textit{PhenomD} waveform generator implemented in the C programming language, but callable from Python. Accompanying this waveform generator is a fast SNR calculator also implemented in C and callable from Python. Both of these C functions are used in the SNR grid generator package. As an additional tool, a Python function combining the \textit{PhenomD} generator and SNR calculator is included in \textit{pyphenomd} for easy calculation of the SNR for binary black holes systems. BOWIE is available for install through the \textit{pip} installation tool. Installation instructions, documentation, and examples within Jupyter Notebooks, including code for producing the plots in this paper, are available at \url{https://github.com/mikekatz04/BOWIE}.

\subsection{Low-Frequency Band Edge Analysis}

\subsubsection{LISA Configurations}
    In our analysis of various LISA configurations, we restrict ourselves to a specific set of sensitivity behaviors. For the purposes of this study, we parameterize the LISA sensitivity in terms of 5 parameters: position noise ($S_p$), acceleration noise ($S_a$), armlength ($A$), break frequency ($f_r$), and low-frequency spectral index ($R$). The break frequency represents the frequency at which the sensitivity curve changes its power law behavior in the low-frequency regime. The exponent in this low-frequency power law is denoted the spectral index. The spectral index is determined using the noise power spectral density (PSD) version of the sensitivity curve. Generally, the break frequency and spectral index are related to the stability of the LISA spacecrafts over long times. At the longest of timescales, affecting the measurements of the largest black holes, the break frequency, spectral index, and acceleration noise will dominate the sensitivity structure at the low-frequency band edge. The armlength is the main contributor to the high-frequency structure of the sensitivity curve; however, longer armlengths do allow for better sensitivity in the low-frequency area of the sensitivity curve, but do not affect the actual upward sloping band edge shape. The position noise sets the noise floor, therefore, determining the most sensitive area of the sensitivity curve. In the LISA development process, engineering teams will work to design LISA to the specifications set forth by the LISA Consortium. For this reason, this paper aims to analyse design parameters from an astrophysical perspective. For more information on the actual design of the spacecraft, please see \cite{LISAMissionProposal}.   
    
\begin{figure}
\begin{center}
\includegraphics[scale=0.36]{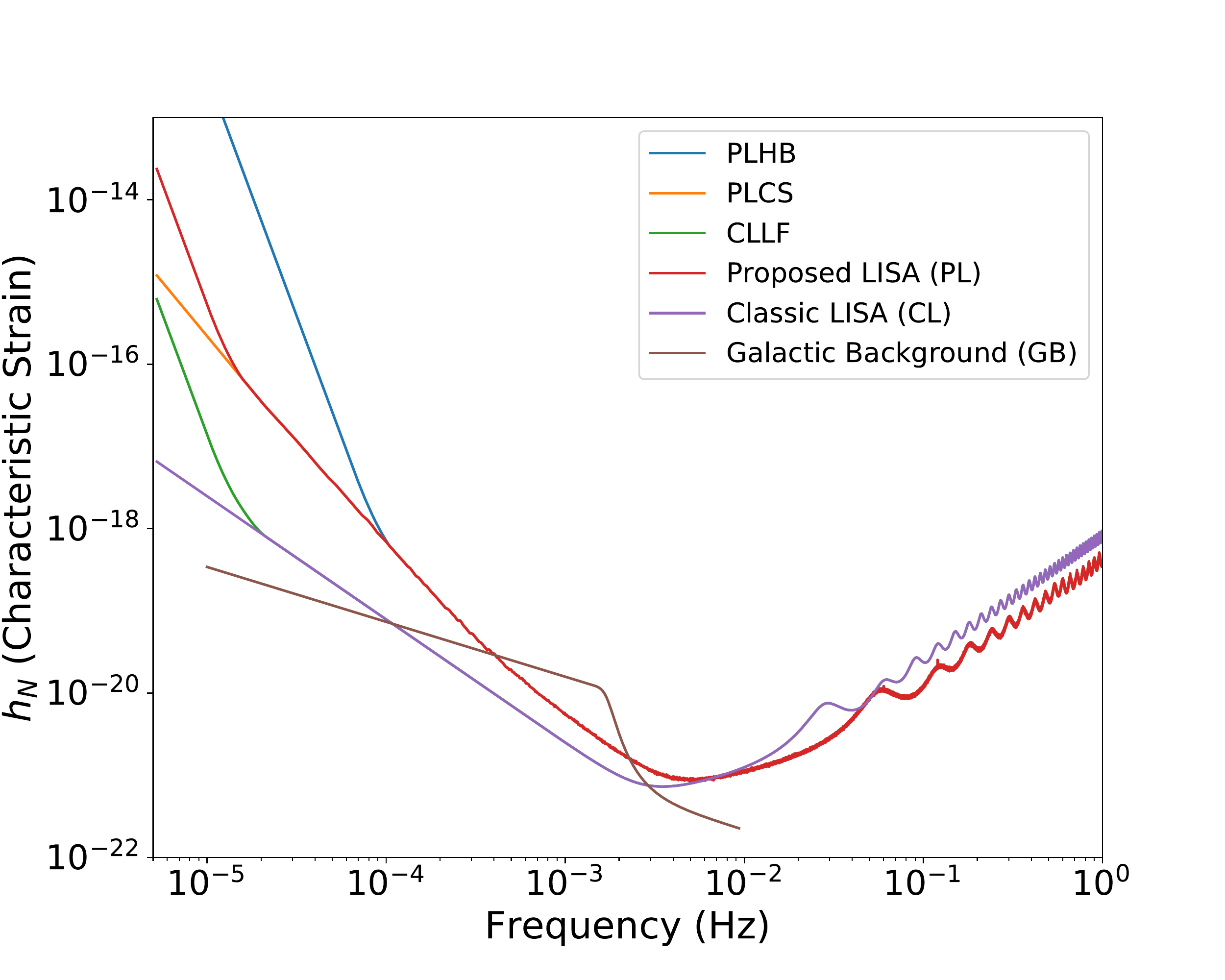}
\end{center}
\caption{All of the LISA sensitivity curves analysed in this paper are shown in terms of the characteristic strain of the noise, $h_N$. The Galactic background is also shown.}
\label{fig:sensecurves-all}
\end{figure}

In this paper, we consider the ``Classic'' \citep{Larson2000} and ``Proposed'' \citep{LISAMissionProposal} LISA configurations to be our base curves from which we adjust, referred to as CL and PL, respectively. All of the sensitivity curves analysed and the additional noise contribution of the white dwarf (WD) background can be seen in figure \ref{fig:sensecurves-all}. In simple terms, CL has better low-frequency performance attributed to a longer armlength and lower spectral index; CL also has an (unrealistic) constant $f^{-2}$ noise performance to infinitely low frequencies, which means it does not contain a specific break frequency.  PL has a break frequency of $\sim1-2\times10^{-5}$Hz and a spectral index of $\sim6$. PL also has slightly better high-frequency performance compared to CL.\footnote{Recently, during LISA phase A studies, the LISA Consortium has adopted a new sensitivity curve available at the LISA project site (\url{cosmo.esa.int/lisa/}). It has a similar high-frequency behavior to CL, while the low-frequency behavior is the same as PL down to $\sim2\times10^{-5}$Hz. As mentioned previously, this frequency acts as the break frequency for PL. Since our study focuses on the low-frequency behavior, we decided to continue using PL. However, this new curve, referred to as LPA for LISA Phase A, and its url address are provided with the BOWIE package.} In order to illustrate the continuum of science performance and how it depends on detector design, we also include analysis with three curves that have different performance characteristics that span a range of possible sensitivity curves between CL and PL. The first curve is a flattened version of the PL sensitivity curve, where the low-frequency spectral index is held constant with no break. This curve is referred to as Proposed LISA constant slope (PLCS). PL is then adjusted by shifting the low-frequency behavior of PL to a higher break frequency at $10^{-4}$ Hz. This curve is denoted Proposed LISA higher break (PLHB). The final curve analysed is constructed by shifting the low-frequency behavior of PL to smaller strains and splining it together with the CL curve, therefore, creating a CL curve with the PL low-frequency behavior. This final curve is identified as the Classic LISA low frequency (CLLF) configuration.
    
    To provide a conservative estimate of the WD background, we use the analytical approximation found in \citet{Hiscock2000} for the background estimation by \citet{HillsBender1997}. As our analysis will show, strong WD background noise will have more influence on the Classic-based curves than the Proposed-based curves since the Classic-based curves have a lower noise characteristic in the area of the Galactic background. The effect of the WD background can be greatly reduced by extended observations and improved global fitting methods as the LISA mission progresses \citep{Cornish2018}.
    
    LISA sensitivity curves are commonly plotted in terms of the square root of the power spectral density, $S_N$. In this study, it is more convenient to use the characteristic strain. The characteristic strain is a representation of the dimensionless strain that includes the effect of integrating a gravitational wave signal over time \citep{Finn2000}. The power spectral density of the noise was converted to characteristic strain of the noise, $h_N$, using $h_N=\sqrt{f S_N}$ \citep{Moore2015}.
  
\subsubsection{Binary Parameter Space}
    In order to study our problem in its entirety, we chose an all-inclusive binary parameter space to examine. All of the following limits are presented to the reader as values measured in the binary's source frame; LISA will measure the redshifted values of the binary's source frame parameters. Our lower limits on the mass were chosen to encompass values of MBH seed binaries that are expected to be astrophysically relevant and detectable by LISA. This study does not consider Galactic or local compact binaries; we analyse black holes larger than $10^2 M_\odot$. The upper limit, $10^{10} M_\odot$, was chosen to represent the largest MBHs believed to exist and/or which have been measured \citep{McConnell2011}. Above $10^{10}M_\odot$, binaries become undetectable by any LISA configuration analysed in this study. Mass ratios between 1 and 500 were considered ($q\geq1$). We examined redshifts as high as 100 and as low as 0.01. The parameter describing the dimensionless spin is given by,
\begin{equation}
	a=\frac{cS}{GM^2},
\end{equation} where $S$ is the magnitude of the spin angular momentum. We only consider spin vectors parallel to the orbital angular momentum. For this parameter, $a$, we examine a range of -0.99 to 0.99. Positive (negative) spins indicate alignment (anti-alignment) with the orbital angular momentum. Due to the qualities of the \textit{PhenomD} model accuracy (see the end of section \ref{sec:charstrain}), we treat both black holes with the same spin parameter (i.e. $a_1 = a_2 = a$). Additionally, we chose to use aligned spins for convenient use of \textit{PhenomD}. Specific to the inspiral stage, we must consider the observation time of LISA. For simplicity, we examine $T_{obs} = 1$ year.

\subsubsection{Characteristic Strain}\label{sec:charstrain}

The dimensionless strain, given by $h(t)$, radiated by a given source represents the change in length between two test masses in a gravitational wave detector, $\Delta L$, normalized by the detector's initial length, $L_0$. To find the characteristic strain radiating from the binary, $h_c$, we used a phenomenological waveform for nonprecessing binaries, \textit{PhenomD} (\citeauthor{Khan2016} \citeyear{Khan2016}; \citeauthor{Husa2016} \citeyear{Husa2016}). \textit{PhenomD} outputs the Fourier transform of the time domain signal, $\tilde{h}(f_s)= F\{h(t)\}(f_s)$, where $f_s$ is the gravitational wave frequency in the source reference frame of the binary. Anything that follows denoted with a subscript $s$ represents a quantity from the binary's source reference frame. This frequency is related to the frequency observed by the LISA detector through the redshift, $z$, of the source: $f_{\rm{obs}} = f_s/(1+z)$. $h_c$ is determined from $\tilde{h}(f_s)$ by \citep{Moore2015}
\begin{equation}\label{eq:hchar}
    \left[h_c(f_s)\right]^2 = 4f_s^2\left|\tilde{h}(f_s)\right|^2.
\end{equation}

With \textit{PhenomD}, we determine the characteristic strain signal using the following source frame inputs: total mass ($M = M_1 + M_2$), symmetric mass ratio ($\eta = M_1 M_2/M^2$), redshift ($z$), dimensionless spin parameter of each black hole ($\chi_i$), and an initial frequency ($f_{st}$) (\textit{st} indicates the start). The luminosity distance ($D_L(z)$) is determined from the redshift using a Planck 2015 cosmology with $H_0 = 67.74$ km/s/Mpc, $\Omega_M = 0.3075$, $\Omega_b=0.0486$, and $\Omega_{vac} = 0.6925$ \citep{Planck2015}. 

The dimensionless spin parameter, $\chi_i$, is given by 
\begin{equation}\label{eq:chi}
    \chi_i = \frac{c}{G}\frac{\vec{S}_i\cdot\hat{L}}{M_i^2},
\end{equation}
where $\vec{S}_i$ is the spin angular momentum vector of the black hole and $\hat{L}$ is the direction of the orbital angular momentum vector. Although $\chi$ can theoretically range from -1.0 through 1.0, we restrict our study to -0.99 to 0.99 on astrophysical grounds. Positive (negative) values represent spins aligned (anti-aligned) with $\hat{L}$. This parameter is equivalent to $a$ for spin vectors parallel or antiparallel to the orbital angular momentum vector, which we use in this paper.

The starting frequency tested for each binary is the gravitational wave frequency 1 year of observation prior to the merger. The merger frequency in the source frame is given by,
\begin{equation}\label{eq:mergerfreq}
   f_{merg,s} = \frac{c^2}{G}\frac{1}{6^{3/2}\pi M_s}.
\end{equation}
Assuming the binary is in a circular orbit, we determine the gravitational wave frequency associated with $T=1$ year of observation prior to the merger, $f_{st}$, by
\begin{equation}
	f_{st,s} = \frac{c^2}{G}\frac{1}{8\pi M_s \tau^{3/8}}\left(1+\left(\frac{11}{32}\eta+\frac{743}{2688}\right)\tau^{-1/4}+...\right),
\end{equation}
where $\tau$ is the dimensionless quantity given by
\begin{equation}\label{eq:tau}
	\tau(T) = \frac{c^3}{G}\frac{\eta T}{5M_s(1+z)}.
\end{equation}
Here, $f_{st,s}$ is quoted to 1 PN order. $T$ of 1 year is given in the reference frame of the detector, and, therefore, must be blue shifted back to a quantity in the source frame of the binary in order to determine $f_{st,s}$ for the \textit{PhenomD} waveform, hence the $1+z$ term in eq. \ref{eq:tau}. This indicates as a source is moved to greater redshifts, the signal received by the detector will represent a time of $1\ \text{year}/(1+z)$ prior to merger.

\begin{figure}
\begin{center}
\includegraphics[scale=0.43]{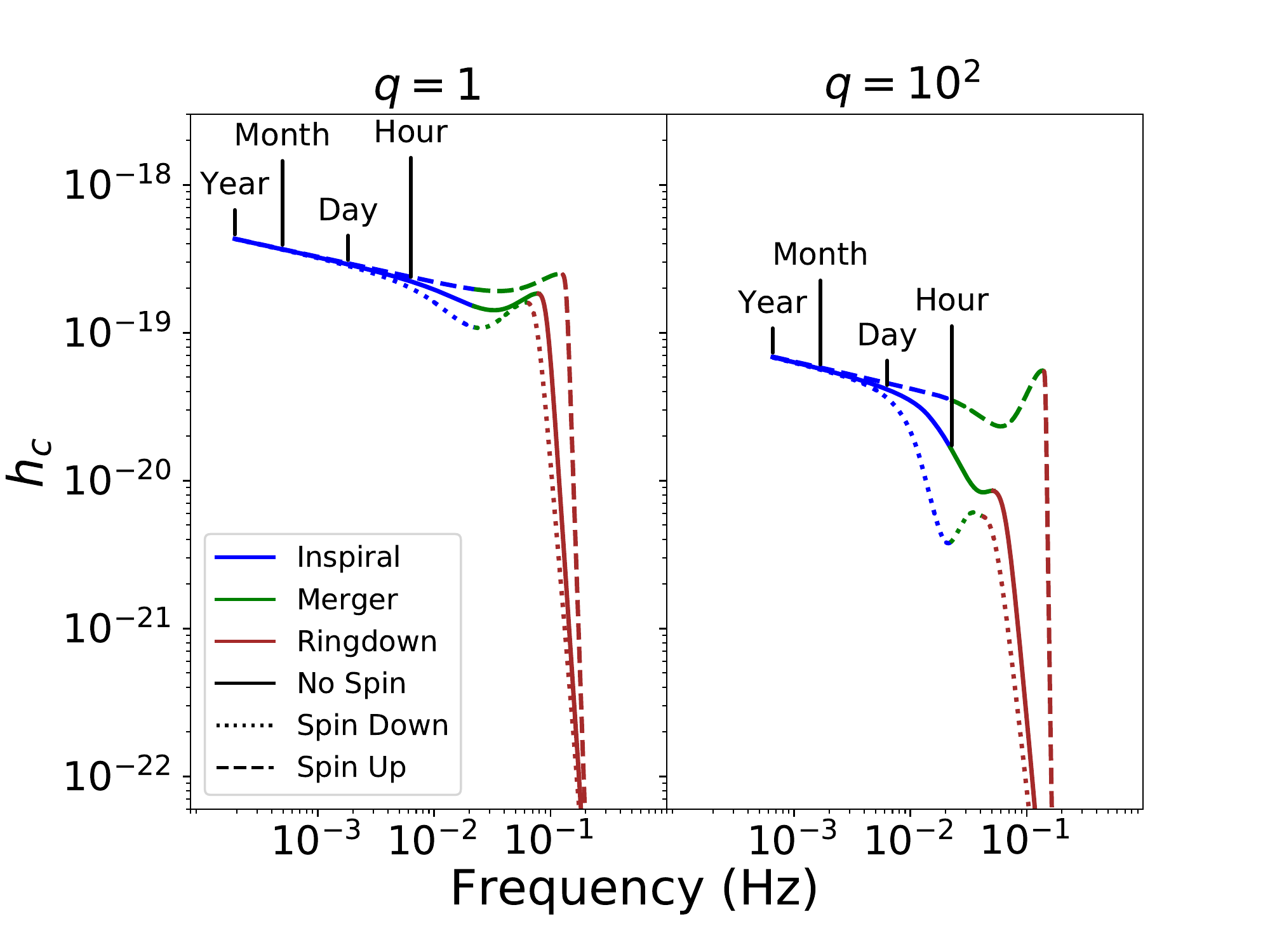}
\end{center}
\caption{The two plots above show comparisons of $h_{c}$  for a binary with a total mass of $10^5\ M_\odot$ at $z=1$ in spin configurations of $a$ equal to -0.99, 0, and 0.99. Mass ratios of $q=1$ and $q=10^{2}$ are shown on the left and right, respectively. Inspiral, merger, and ringdown stages are shown in blue, green, and brown, respectively. The times before the completion of the merger are labelled for reference.}
\label{fig:hcharexamples}
\end{figure}

Ringdown begins at the frequency of the peak that occurs after merger. In the formalism of \textit{PhenomD}, this peak value is the frequency of the $l=m=2$ quasinormal ringing mode of the final black hole shifted due to a damping factor. See \citet{Khan2016} for details on this construction. Figure \ref{fig:hcharexamples} shows examples of $h_c$ deciphering between the three stages of binary black hole coalescence for different $q$ and $a$. In general, moving from $a=-0.99$ to $a=0.99$ will increase $h_c$. However, this disparity increases as the mass ratio becomes farther from unity. 

\textit{PhenomD} is calibrated using numerical relativity waveforms up to $q=18$ with spins up to $a\sim0.85$. It is calibrated up to $a\sim0.98$ for equal-mass systems. \textit{PhenomD} shows mismatches of less than $\sim1\%$ in its calibration region. \citet{Khan2016} state that outside its calibration region, \textit{PhenomD} produces physically reasonable results. However, they express caution in assuming the late-stage merger and ringdown waveforms are entirely accurate for applications where high fidelity waveforms are needed (e.g. parameter estimation). With that in mind, the authors state that \textit{PhenomD} ``does not show any pathological behavior outside its calibration region, neither in the time nor frequency domain.'' Even though this does not guarantee accuracy, the model is more robust in terms of extrapolation. In terms of spins, the model is designed around a reduced-spin approximation where an effective spin parameter represents a specific mass-weighted combination of the two black hole spins. The authors expect for this approximation to be less accurate for higher mass ratios and high aligned spins. Additionally, the model was mostly calibrated using equal-spin numerical relativity waveforms. This is our motivation for using the same spin for both black holes, $a_1=a_2=a$. See section IX of \citet{Khan2016} for more details on the accuracy of \textit{PhenomD}.

\subsubsection{Signal-to-Noise Ratio}

With $h_c$ and $h_N$ determined, the SNR of a coalescence signal can be found by integrating the ratio of $h_c$ to $h_N$ over frequency space. The SNR for LISA averaged over sky location, polarization, and inclination is given by \citep{Cornish2018},
\begin{equation}\label{eq:SNR}
    \rho^2 = \frac{64}{5}\int_0^\infty \frac{\left|\tilde{h}(f)\right|^2}{S_N(f)}df = \frac{16}{5}\int_0^\infty\frac{1}{f}\frac{h_c^2(f)}{h_N^2(f)}df.
\end{equation}
The SNR is then multiplied by a factor of $\sqrt{2}$ because we consider a 2-channel detector. The SNR is used as a tool to determine LISA's detectability. See \citet{Robson2017} and sources within for more details on LISA parameter estimation. 

Additionally, we comment on the detectability of observable sources resolvable for the purposes of ``black hole spectroscopy.'' This practice involves measuring subdominant modes in the ringodwn spectrum. \citet{Berti2016} lay out a foundation for using the SNR to determine detectability of these subdominant modes.  These measurements can be used to address discrepancies in general relativity and the Kerr metric. The General Likelihood Ratio Test (GLRT) is used to determine if either the $l=m=3$ or $l=m=4$ modes are resolvable. The SNRs from the dominant $l=m=2$ mode necessary for detecting these subdominant modes are given by
\begin{equation}\label{eq:spec3}
    \rho_{GLRT}^{2,3} = 17.687 + \frac{15.4597}{q-1}-\frac{1.65242}{q},
\end{equation}
\begin{equation}\label{eq:spec4}
\rho_{GLRT}^{2,4} = 37.9181 + \frac{83.5778}{q}+\frac{44.1125}{q^2}+\frac{50.1316}{q^3}.
\end{equation}
Spectroscopic measurements can be performed if $\rho$ is greater than $\rho_{GLRT}\equiv\text{min}\left(\rho_{GLRT}^{2,3},\rho_{GLRT}^{2,4}\right)$ \citep{Berti2016}.

\subsubsection{Signals Near the Band-Edge}

\begin{figure}
\begin{center}
\includegraphics[scale=0.36]{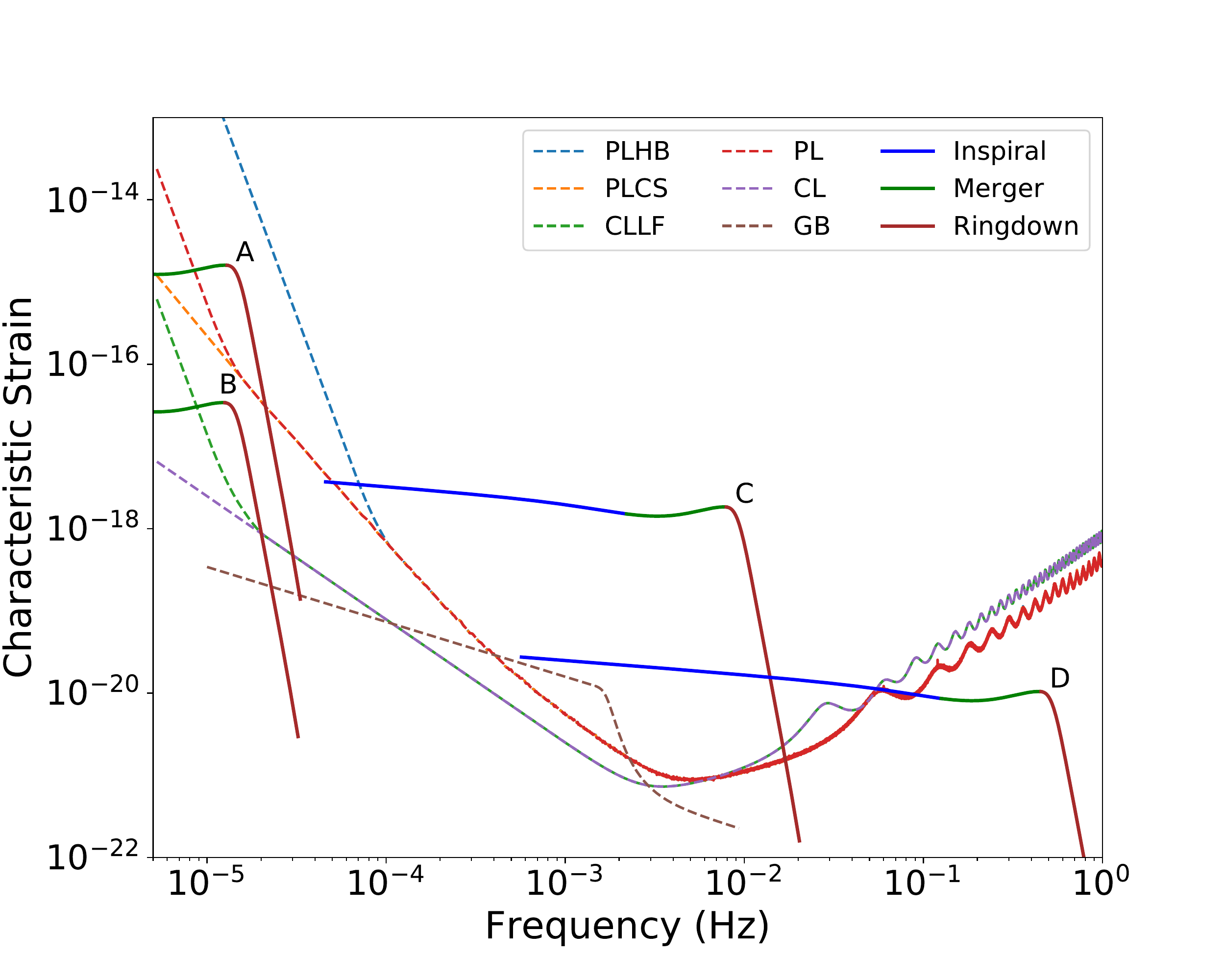}
\end{center}
\caption{Four examples of $h_c$ are shown and how its relation to $h_N$ can affect the SNR measured by LISA. The binaries shown are in the $a=0$ and $q=1$ configuration with the following parameters: (A) $M= 7\times10^8M_\odot$ at $z=0.75$ (B) $M = 6\times10^7M_\odot$ at $z=20$ (C) $M = 1\times10^6M_\odot$ at $z=1.0$ (D) $M= 1\times10^4M_\odot$ at $z=2.5$.} 
\label{fig:hcversushn}
\end{figure}

Figure \ref{fig:hcversushn} provides four examples of how the comparison of $h_c$ to $h_N$ affects measurements by LISA. Detectable sources can be approximated using characteristic strain curves by observing which source curves lie above each sensitivity curve. The SNR can be roughly compared between source and sensitivity curve configurations by integrating the area between $h_c$ and $h_N$ ``by eye.'' All of the examples are shown in the $a=0$ and $q=1$ configuration and were chosen for the purpose of illustrating sensitivity differences. The overall signal SNR, with the inclusion of the WD background, for each example binary and sensitivity curve combination is shown in table \ref{tb:SNR_for_examples}. Example A shows a binary with $M = 7\times10^8M_\odot$ at $z=0.75$. Both CL and CLLF will make measurements of the end of merger as well as a good measurement of the ringdown. Since most of the rindown SNR is accumulated near its peak \citep{FlanaganHughes1}, PL and PLCS will measure the ringdown of this binary, but only PLCS will be able to read any signal emanating from the merger. The curve with the steeper break, PLHB, will not detect this binary. The next binary, modeled as example B, represents $M = 6\times10^7M_\odot$ at $z=20$. This binary is chosen to illustrate the difference between the PL and CL base curves. None of the curves whose base is PL will be able to detect this binary. However, both curves with a base of CL will measure this binary's merger and ringdown. If we would have tested a curve with the low-frequency behavior of PLHB attached to a base of CL, this binary would not be visible. These first two example binaries represent the area where the LISA band edge is critical to detecting and analyzing binaries in the larger mass range. Example C displays a prime target for the LISA mission: a binary with $M = 1\times10^6M_\odot$ at $z=1.0$. This is a prime target due to its time spent evolving during inspiral within the LISA band in addition to a full measurement of its merger and ringdown. For the inspiral stage, CL and CLLF will make slightly stronger measurements compared to Proposed-based curves. However, the WD background does negate much of the differences between sensitivity curves in this region. The final binary shown, example D, represents a binary of IMBHs with $M = 1\times10^4M_\odot$ at $z=2.5$. This binary will be detectable only during its inspiral by all sensitivity curves analysed here. With the addition of the WD background, the SNRs for all sensitivity curves are within a small factor of each other for this binary.

\begin{table}
\centering
 \begin{tabular}{||c c c c c c||} 
 \hline
Example & PL & CL & PLHB & PLCS & CLLF \\ [0.5ex] 
 \hline
A & 27 & 2386 & 0 & 34 & 1072 \\
B & 0 & 48 & 0 & 0 & 20 \\
C & 7274 & 7984 & 7274 & 7274 & 7984 \\
D & 96 & 103 & 96 & 96 & 103 \\
 \hline
\end{tabular}
\caption{Overall SNR values for each example binary and sensitivity curve combination shown in figure \ref{fig:hcversushn}. The WD background is included.}
\label{tb:SNR_for_examples}
\end{table}


\section{Results}

\subsection{Overall Signal Comparison}

\begin{figure*}
\begin{center}
\includegraphics[scale=0.55]{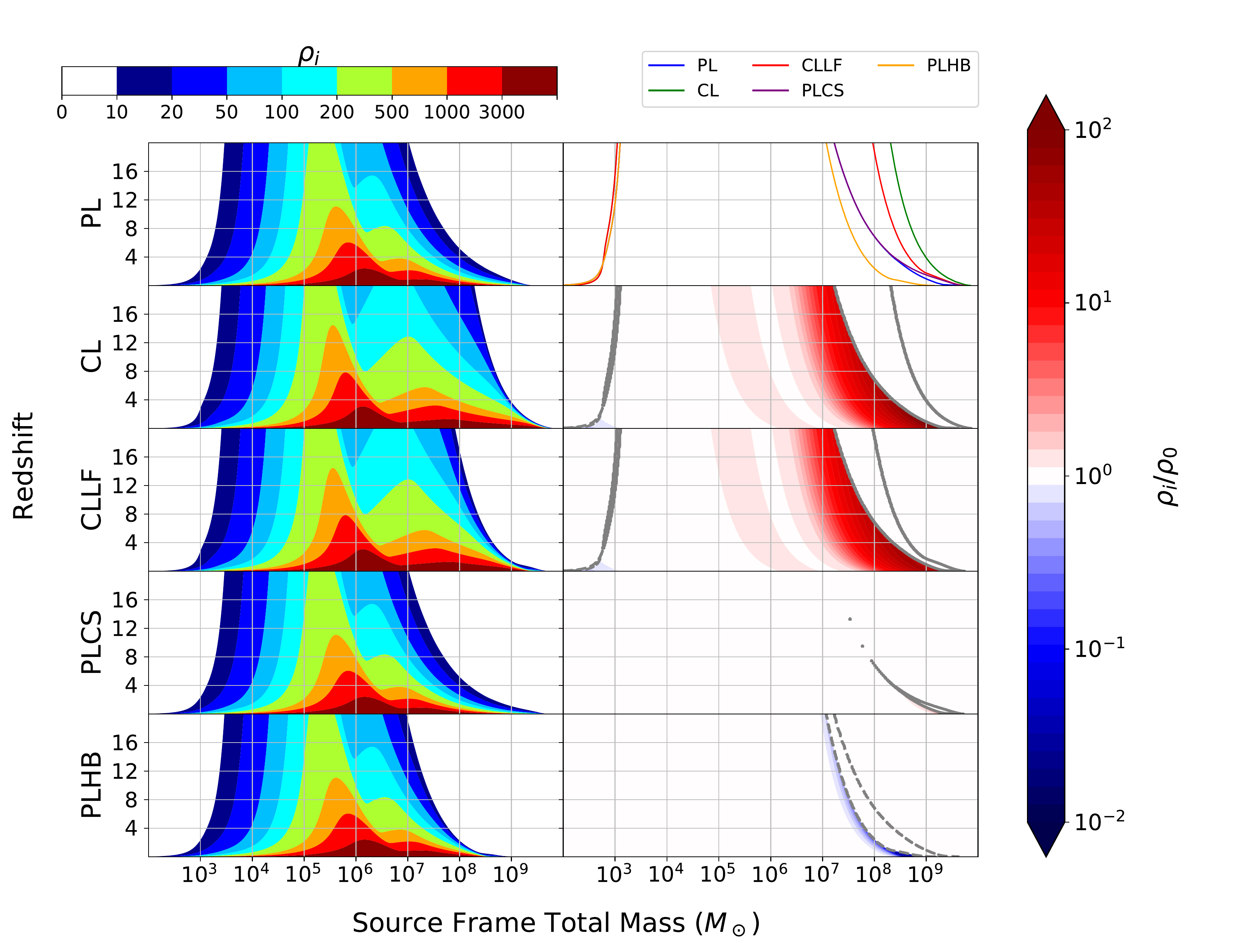}
\end{center}
\caption{This plot shows an output of BOWIE comparing a baseline configuration (PL, top row) against alternative configurations (all other rows). All binaries are shown with $q=5$ and $a=0.8$, including the Galactic background noise. The source frame total mass is plotted against the redshift. The left column represents waterfall plots according to the top color bar. The top of the right column shows the horizon plots of all curves analysed for $\rho=5$. The rest of the right column contains SNR ratio plots with PL as the baseline ($\rho_0$), according to the right color bar. Ratio contours show $\rho_i/\rho_0$, where $\rho_i$ is the SNR from the alternative configuration in each row. The ratio is only shown if $\rho_i>5$ and $\rho_0>5$. Loss/gain contours are overlaid on top of the ratio plots. A gain contour ($\rho_i>5$, $\rho_0<5$) outlines the parameter space with a solid gray line where sources can be detected by the alternative but not the baseline configuration. A loss contour ($\rho_i<5$, $\rho_0$>5) outlines the parameter space with a dashed gray line where sources can be detected by the baseline but not the alternative configuration. Loss/gain contours are not plotted for sources undetectable by either configuration ($\rho_i<5$, $\rho_0<5$), nor for sources detectable by both configurations ($\rho_i>5$, $\rho_0>5$).} 
\label{fig:maincomparisonWD}

\end{figure*}

The first comparison represents the SNR from the entire signal, including inspiral, merger, and ringdown, beginning at one year before reaching merger. This comparison  can be seen in figure \ref{fig:maincomparisonWD}. The left column represents waterfall plots for each sensitivity curve, labelled alongside the y-axis, corresponding to the top color bar shown. The top row, indicating the PL configuration, will be our baseline against which we compare the other sensitivity curves, designated $\rho_0$. The top of the right column is a horizon plot representing the redshift horizon of each configuration based on our cut of $\rho=5$. The remainder of the right column displays ratio plots of $\rho_i/\rho_0$, where $\rho_i$ is the SNR of the alternative sensitivity curve in each row. The ratio plots correspond to the color bar on the right. The red (blue) contours display an area where the comparison curve has better (worse) performance than PL. The ratio contours are only shown when the source is detectable by both configurations ($\rho_i>5$ and $\rho_0$>5). Overlaid on these plots are loss/gain contours shown in gray. A gain contour ($\rho_i>5$, $\rho_0<5$) outlines the parameter space with a solid gray line where sources can be detected by the alternative but not the baseline configuration. A loss contour ($\rho_i<5$, $\rho_0$>5) outlines the parameter space with a dashed gray line where sources can be detected by the baseline but not the alternative configuration. Loss/gain contours are not plotted for sources undetectable by either configuration ($\rho_i<5$, $\rho_0<5$), nor for sources detectable by both configurations ($\rho_i>5$, $\rho_0>5$). Here, we examine binaries with a mass ratio $q=5$ and spin $a=0.8$. The mass ratio was chosen to represent strong signal binaries that are close to but not the same as the equal mass case. Astrophysical measurements of central nuclei black holes have shown spin near the maximum Kerr value (\citeauthor{Miller2007} \citeyear{Miller2007}; \citeauthor{Reis2008} \citeyear{Reis2008}). We do a separate spin comparison across the range of aligned spin values, showing $a$ values of -0.99, 0.0, and 0.99. Therefore, we chose $a=0.8$ to show a high spin value that will differ from the $a=0.99$ comparison.

Depending on the low-frequency behavior of the band edge, this signal can diminish quicker or slower as we move to higher total masses. As expected, the CL configuration has better low-frequency performance than PL, allowing for better detection of MBHs at the high-mass end. The measurement strength of the IMBH regime is about the same for CL and PL. PL does have slightly better high-frequency performance, leading to better measurements of near-stellar mass black hole binaries. 

The CLLF configuration behaves similarly to the CL configuration, but has slightly worse performance in the high mass regime due to its break frequency and higher spectral index. Consequently, CLLF measures stronger signals for MBH sources compared to PL, but not quite as strong as CL for above $10^8M_\odot$. In general, the comparison of PL to CLLF can be viewed as how the SNR changes when the low-frequency behavior is the same, but is shifted towards lower noise levels via changes in the armlength or acceleration noise.

When comparing PL to PLCS and PLHB, no considerable difference is seen until just below $\sim10^9M_\odot$. PLCS has about the same measurement capabilities as PL, with a slightly better measurement for only the largest detectable MBHs.\footnote{The loss/gain contour shown for PLCS looks like a dashed contour because it is limited by grid resolution. The line should be entirely solid.} Therefore, the effect of maintaining the base parameters of PL with a constant spectral index and no break frequency is nonzero, but minimal.

The PLHB configuration is the one sensitivity curve displaying worse performance than PL on the high mass end, as expected due to its higher break frequency. Moving the break frequency towards higher frequencies while maintaining the base parameters of PL will diminish the signal of the largest black holes observable by LISA.

The effect of the WD noise is most important to the overall SNRs in the middle of the mass parameter space. This noise affects the area of the sensitivity curves where IMBHs inspiral and lower mass MBHs merge. However, the negative effect is greater on the Classic-based configurations than on the Proposed-based configurations (see figure \ref{fig:sensecurves-all}), therefore, negating their differences in the middle area of the parameter space. At low frequencies, where the largest black holes merge, the WD background plunges below all sensitivity curves. Therefore, the measurement of the largest MBH sources is not greatly affected by the WD background.

\subsection{Comparing Stages of Binary Coalescence}

\begin{figure*}
\begin{center}
\includegraphics[scale=0.55]{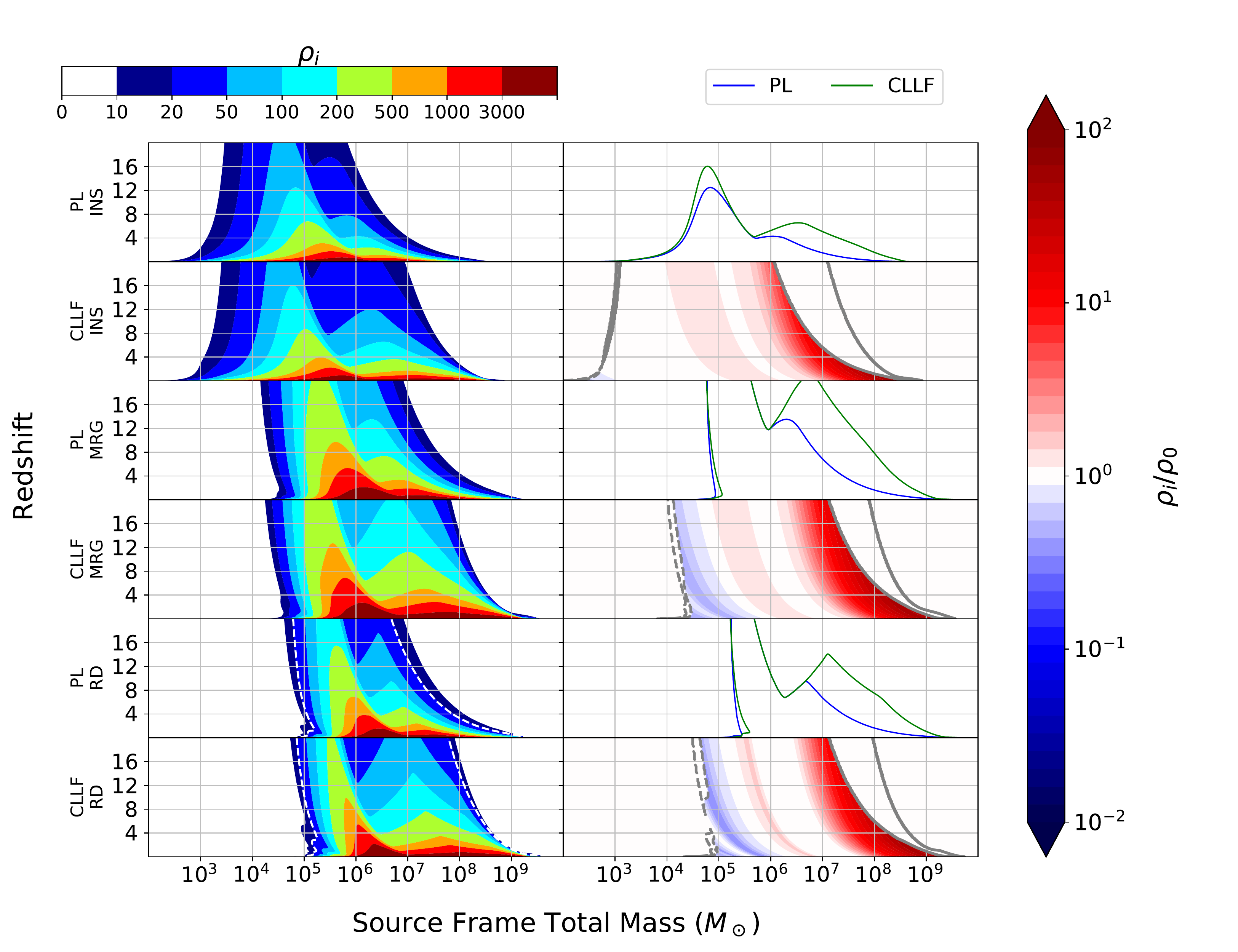}
\end{center}
\caption{A comparison of CLLF to PL is shown for the three stages of binary black hole coalescence. Here, $q=5$ and $a=0.8$. Inspiral (INS), merger (MRG), and ringdown (RD) are shown successively from top to bottom in sets of two rows each (labelled along the vertical axis). The right column contains two different plots: horizon and SNR ratio plots with loss/gain contours separated for inspiral, merger, and ringdown signals. Refer to the caption of figure \ref{fig:maincomparisonWD} for information on the ratio and loss/gain construction. The horizon plots show an SNR contour of $\rho=100$. The ringdown waterfall plots also display a spectroscopic contour for the $l=m=3$ quasinormal mode with a white dashed line. With $q=5$, equation \ref{eq:spec3} determines the minimum SNR for the $l=m=2$ mode to be 21 for spectroscopic measurements.} 
\label{fig:phasecomp}
\end{figure*}

In this section, we provide plots comparing only PL and CLLF in each of the three stages of binary black hole coalescence. Binaries are once again shown in the $q=5$ and $a=0.8$ configuration. This comparison is shown in figure \ref{fig:phasecomp}. One aspect to note here is the loss/gain plots show exactly the difference between the redshift horizons. Since the comparison here is between only two plots, a horizon plot with contours for the SNR cut would be redundant to that shown in the loss/gain contours. Therefore, the horizon plots shown in this comparison display SNR contours of 100. This arbitrary value is chosen to represent the ability of each LISA configuration to strongly (rather than marginally) detect sources. 

The inspiral plots (top two), show virtually the same exact behavior as the main comparison shifted towards lower masses (left) and slightly lower redshifts (down). As seen in figure \ref{fig:hcharexamples}, the inspiral portion of coalescence spans most of the frequency regime representing $h_c$. The low-frequency edge of the inspiral spectrum contacts and moves below $h_N$ at the same place as the overall waveform, leading to similar band edge behaviors.

In the merger comparison, the high mass cutoff is close to but not the same as the overall plots due to the missing extension of the ringdown farther into band. During merger and ringdown, large SNR accumulation occurs near the peak in the frequency spectrum. This feature results in abrupt breaks in ratio contours within the IMBH region, seen in both the merger and ringdown comparisons. These breaks are due to the locations of the high frequency structure in the sensitivity curve, along with the location of the Galactic background. This structure causes the noise level to increase and decrease rapidly on smaller frequency scales, causing the location of the waveform peak to have a heavy influence on the SNR. For this reason, the more minute differences in the structure of the sensitivity curve are magnified. This results in PL having better measurements for IMBHs during the merger, while CLLF maintains its stronger signal in the MBH regime. 

The ringdown signals are shifted slightly to higher masses as it is the highest frequency portion of the waveform and is, therefore, cutoff last by the band edge. The loss of MBH sources by PL is important here since the ringdown may be the only observable portion of the spectrum for the largest MBHs.  

An additional contour (dashed white line) is added to the waterfall plots for the ringdown. These contours represent an SNR of 21 for the $l=m=2$ ringdown mode, which corresponds to $\rho_{GLRT}$ for $q=5$, the SNR necessary for black hole spectroscopy. Here, this is for the $l=m=3$ quasinormal mode. The differences in these contours between PL and CLLF show CLLF will have a better chance at spectroscopic measurements for large MBHs, even out to higher redshifts. 

\subsection{Mass Ratio and Spin Comparison}

\begin{figure*}
\begin{center}
\includegraphics[scale=0.55]{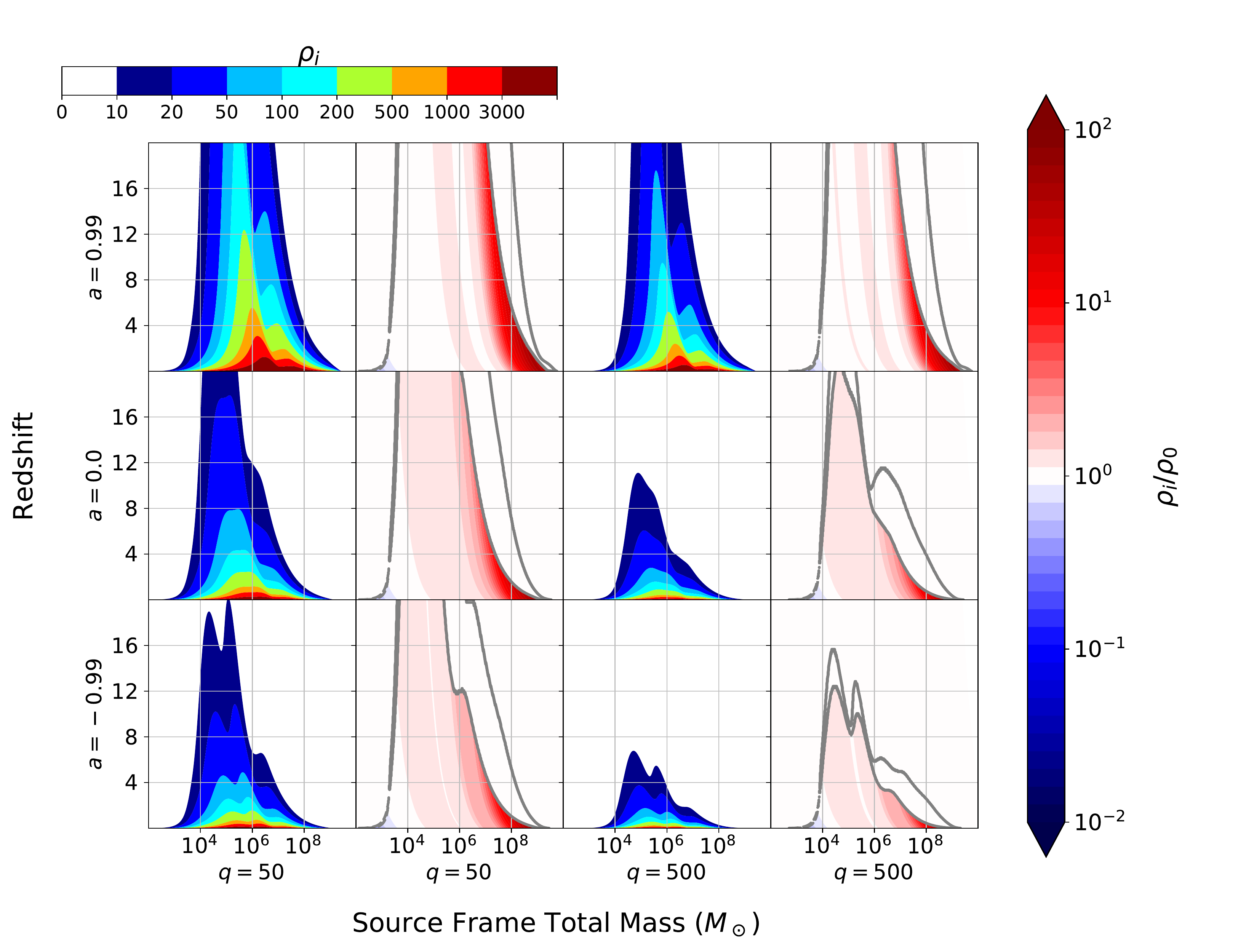}
\end{center}
\caption{These plots show a comparison of CLLF and PL across spin values and higher mass ratios. The first and second columns represent mass ratios of 50, while the third and fourth show mass ratios of 500. The rows, from top to bottom, represent spins of 0.99, 0.0, and -0.99. A waterfall plot is shown for each combination for a PL sensitivity curve. Each waterfall plot has a corresponding ratio plot with loss/gain contours, comparing CLLF (alternative) to PL (baseline), shown directly to its right. Refer to the caption for figure \ref{fig:maincomparisonWD} for information on the ratio and loss/gain construction.}
\label{fig:massratiospincompWD}
\end{figure*}

Figure \ref{fig:massratiospincompWD} shows a comparison for the entire signal from gridded binary configurations with mass ratios of 50 and 500 and spins of 0.99, 0.0, and -0.99 (at $q=5$, the spin does not change the signal significantly). The corresponding rows and columns for each spin and mass ratio configuration are labelled along the left and bottom axes, respectively. The figure is grouped into two sets separated by the central vertical line: $q=50$ binaries are shown in the first two columns and $q=500$ binaries are shown in the last two columns. In each set, waterfall plots for each spin value, with a PL sensitivity curve, are shown in the left column. Each waterfall plot has a corresponding ratio plot, comparing CLLF (alternative) to PL (baseline) as in figure \ref{fig:phasecomp}, which is shown to the right of each waterfall plot.

For all plots shown, CLLF makes a stronger measurement on the largest MBH binaries. As $q$ increases and/or $a$ decreases, the signal diminishes quickly, making high-mass fringe detections more dependent on the sensitivity.\footnote{Please note at higher mass ratios, the binary radiates more energy to higher quasinormal modes during ringdown.} This is important for many realistic candidates for MBH mergers with a large seed black hole at $\sim10^5M_\odot$ or a Milky Way-like MBH at $\sim10^6M_\odot$ spiraling into a $\sim10^8M_\odot$ MBH at the centre of a larger galaxy. These types of mergers are commonly seen in cosmological simulations like Illustris (\citeauthor{Blecha2016} \citeyear{Blecha2016}, \citeauthor{PTA-illustris} \citeyear{PTA-illustris}).

\section{Conclusions}

Using the latest phenomenological waveforms, we computed the SNR for an array of potential binary black hole systems with five different sensitivity curves. Each curve tested shows one effective major difference to Proposed LISA. We focused on the low-frequency band edge behavior of these curves as this area of the frequency spectrum directly affects the measurement of MBH binaries resulting from galaxy mergers. Ultimately, the low-frequency performance of LISA has important ramifications on its ability to probe the astrophysical population of large MBH binaries with total masses of $\sim10^7M_\odot-10^9M_\odot$. We found the most important factor for MBH detectability is the armlength and acceleration noise: curves similar to Classic LISA will provide the best measurements of the largest MBHs. The break frequency is the second most impactful quantity as a high break frequency will effectively cut off the largest MBHs entirely. The spectral index will affect overall measurement capability, but its impact will be small compared to the other two quantities, assuming it is reasonably physical, rather than directly vertical.

We tested the response of each detector to the stages of binary coalescence. This proved interesting for merger and ringdown due to their high signal gain near the peak in the characteristic strain spectrum (see figure \ref{fig:phasecomp}). MBH signals are stronger with the Classic LISA low frequency configuration while IMBH signals generally remain stronger with Proposed LISA due to the Classic LISA low frequency configuration's better low-frequency performance and Proposed LISA's better high-frequency performance. The Classic LISA low frequency configuration does have better capability of measuring spectroscopic sources, especially at higher masses.

At higher mass ratios and lower spins, the band edge shape gains importance for its ability to bring in fringe sources on the high mass end. Therefore, a configuration similar to Classic LISA will open up the parameter space containing these threshold binaries. If most high mass ratio sources turn out to be closer to the spin up configuration, a much larger parameter space coverage is attainable.

For our analysis we developed a new Python tool for representing the effect of LISA sensitivity curve changes on LISA's scientific impact. We call this tool Binary Observability With Illustrative Exploration, or BOWIE. It is publicly available on github at \url{https://github.com/mikekatz04/BOWIE}. It includes three Python packages that are extensible to questions beyond those addressed here. The first package, \textit{pyphenomd}, is a \textit{PhenomD} waveform generator and SNR calculator for binary black hole gravitational wave analysis. The second package, \textit{bowie\_gencondata}, uses the waveform generator package to create SNR grids across a user chosen parameter space. These grids are used as input for the third package, \textit{bowie\_makeplot}, to create plots like those shown in this paper. 

Our analysis shows the difference between each sensitivity curve's response to our binary parameter space using different visualizations of SNR information. Waterfall plots show the SNR from one sensitivity curve and binary configuration as a filled contour plot. Horizon plots show line contours at a singular SNR value for many configurations. In this paper, we introduced two new  graphical representations for comparing the SNR from various sensitivity configurations: ratio plots and loss/gain contours. Ratio plots show contours of the ratio between SNRs of two configurations where both configurations are above the detectable SNR limit. Loss/gain contours show where the two configurations differ in terms of detectable sources, indicating sources which meet an SNR threshold with one configuration but not the other. Together, ratio plots and loss/gain contours reveal pertinent information about the impact of configuration changes on LISA science in a meaningful and understandable form.  

\section*{Acknowledgements}
This research was supported by the National Science Foundation under grant DGE-1450006 and the Illinois Space Grant Consortium.

MLK would like to thank Scott Coughlin for helpful discussion on frequency domain waveform creation and Brandon Miller for general help in the initial stages of this project. 

Astropy, a community-developed core Python package for Astronomy, was used in this research \citep{Astropy}. This paper also employed use of Scipy \citep{scipy}, Numpy \citep{Numpy}, and Matplotlib \citep{Matplotlib}.



\pagebreak
\bibliographystyle{mnras}
\bibliography{bibliography} 

\begin{thebibliography}{}
\makeatletter
\relax
\def\mn@urlcharsother{\let\do\@makeother \do\$\do\&\do\#\do\^\do\_\do\%\do\~}
\def\mn@doi{\begingroup\mn@urlcharsother \@ifnextchar [ {\mn@doi@}
  {\mn@doi@[]}}
\def\mn@doi@[#1]#2{\def\@tempa{#1}\ifx\@tempa\@empty \href
  {http://dx.doi.org/#2} {doi:#2}\else \href {http://dx.doi.org/#2} {#1}\fi
  \endgroup}
\def\mn@eprint#1#2{\mn@eprint@#1:#2::\@nil}
\def\mn@eprint@arXiv#1{\href {http://arxiv.org/abs/#1} {{\tt arXiv:#1}}}
\def\mn@eprint@dblp#1{\href {http://dblp.uni-trier.de/rec/bibtex/#1.xml}
  {dblp:#1}}
\def\mn@eprint@#1:#2:#3:#4\@nil{\def\@tempa {#1}\def\@tempb {#2}\def\@tempc
  {#3}\ifx \@tempc \@empty \let \@tempc \@tempb \let \@tempb \@tempa \fi \ifx
  \@tempb \@empty \def\@tempb {arXiv}\fi \@ifundefined
  {mn@eprint@\@tempb}{\@tempb:\@tempc}{\expandafter \expandafter \csname
  mn@eprint@\@tempb\endcsname \expandafter{\@tempc}}}

\bibitem[\protect\citeauthoryear{{Amaro-Seoane} et~al.,}{{Amaro-Seoane}
  et~al.}{2017}]{LISAMissionProposal}
{Amaro-Seoane} P.,  et~al., 2017, preprint, \href
  {http://adsabs.harvard.edu/abs/2017arXiv170200786A} {} (\mn@eprint {arXiv}
  {1702.00786})

\bibitem[\protect\citeauthoryear{{Astropy Collaboration} et~al.,}{{Astropy
  Collaboration} et~al.}{2013}]{Astropy}
{Astropy Collaboration} et~al., 2013, \mn@doi [\aap]
  {10.1051/0004-6361/201322068}, \href
  {http://adsabs.harvard.edu/abs/2013A%26A...558A..33A} {558, A33}

\bibitem[\protect\citeauthoryear{{Baibhav}, {Berti}, {Cardoso}  \&
  {Khanna}}{{Baibhav} et~al.}{2018}]{Baibhav2018}
{Baibhav} V.,  {Berti} E.,  {Cardoso} V.,   {Khanna} G.,  2018, \mn@doi [\prd]
  {10.1103/PhysRevD.97.044048}, \href
  {http://adsabs.harvard.edu/abs/2018PhRvD..97d4048B} {97, 044048}

\bibitem[\protect\citeauthoryear{{Bansal}, {Taylor}, {Peck}, {Zavala}  \&
  {Romani}}{{Bansal} et~al.}{2017}]{Bansal2017}
{Bansal} K.,  {Taylor} G.~B.,  {Peck} A.~B.,  {Zavala} R.~T.,   {Romani} R.~W.,
   2017, \mn@doi [\apj] {10.3847/1538-4357/aa74e1}, \href
  {http://adsabs.harvard.edu/abs/2017ApJ...843...14B} {843, 14}

\bibitem[\protect\citeauthoryear{{Barausse}, {Bellovary}, {Berti},
  {Holley-Bockelmann}, {Sathyaprakash}  \& {Sesana}}{{Barausse}
  et~al.}{2015}]{Barausse2015}
{Barausse} E.,  {Bellovary} J.,  {Berti} E.,  {Holley-Bockelmann} K.~{Farris}
  B.,  {Sathyaprakash} B.,   {Sesana} A.,  2015, J. Phys.: Conf. Ser., 610,
  012001

\bibitem[\protect\citeauthoryear{{Bender} \& {Hils}}{{Bender} \&
  {Hils}}{1997}]{HillsBender1997}
{Bender} P.~L.,  {Hils} D.,  1997, \mn@doi [Classical and Quantum Gravity]
  {10.1088/0264-9381/14/6/008}, \href
  {http://adsabs.harvard.edu/abs/1997CQGra..14.1439B} {14, 1439}

\bibitem[\protect\citeauthoryear{{Berti}, {Cardoso}  \& {Will}}{{Berti}
  et~al.}{2006}]{Berti2006}
{Berti} E.,  {Cardoso} V.,   {Will} C.~M.,  2006, \mn@doi [\prd]
  {10.1103/PhysRevD.73.064030}, \href
  {http://adsabs.harvard.edu/abs/2006PhRvD..73f4030B} {73, 064030}

\bibitem[\protect\citeauthoryear{{Berti}, {Sesana}, {Barausse}, {Cardoso}  \&
  {Belczynski}}{{Berti} et~al.}{2016}]{Berti2016}
{Berti} E.,  {Sesana} A.,  {Barausse} E.,  {Cardoso} V.,   {Belczynski} K.,
  2016, \mn@doi [Phys. Rev. Lett.] {10.1103/PhysRevLett.117.101102}, \href
  {http://adsabs.harvard.edu/abs/2016PhRvL.117j1102B} {117, 101102}

\bibitem[\protect\citeauthoryear{{Blecha} et~al.,}{{Blecha}
  et~al.}{2016}]{Blecha2016}
{Blecha} L.,  et~al., 2016, \mn@doi [\mnras] {10.1093/mnras/stv2646}, \href
  {http://adsabs.harvard.edu/abs/2016MNRAS.456..961B} {456, 961}

\bibitem[\protect\citeauthoryear{{Boehle} et~al.,}{{Boehle}
  et~al.}{2016}]{BoehleGhez2016}
{Boehle} A.,  et~al., 2016, \mn@doi [\apj] {10.3847/0004-637X/830/1/17}, \href
  {http://adsabs.harvard.edu/abs/2016ApJ...830...17B} {830, 17}

\bibitem[\protect\citeauthoryear{{Cornish} \& {Robson}}{{Cornish} \&
  {Robson}}{2018}]{Cornish2018}
{Cornish} N.,  {Robson} T.,  2018, preprint, \href
  {http://adsabs.harvard.edu/abs/2018arXiv180301944C} {} (\mn@eprint {arXiv}
  {1803.01944})

\bibitem[\protect\citeauthoryear{{Dosopoulou} \& {Antonini}}{{Dosopoulou} \&
  {Antonini}}{2017}]{Dosopoulou2017}
{Dosopoulou} F.,  {Antonini} F.,  2017, \mn@doi [\apj]
  {10.3847/1538-4357/aa6b58}, \href
  {http://adsabs.harvard.edu/abs/2017ApJ...840...31D} {840, 31}

\bibitem[\protect\citeauthoryear{{Finn} \& {Thorne}}{{Finn} \&
  {Thorne}}{2000}]{Finn2000}
{Finn} L.~S.,  {Thorne} K.~S.,  2000, \mn@doi [\prd]
  {10.1103/PhysRevD.62.124021}, \href
  {http://adsabs.harvard.edu/abs/2000PhRvD..62l4021F} {62, 124021}

\bibitem[\protect\citeauthoryear{Flanagan \& Hughes}{Flanagan \&
  Hughes}{1998}]{FlanaganHughes1}
Flanagan E.~E.,  Hughes S.~A.,  1998, \mn@doi [Phys. Rev. D]
  {10.1103/PhysRevD.57.4535}, 57, 4535

\bibitem[\protect\citeauthoryear{{Gair}, {Vallisneri}, {Larson}  \&
  {Baker}}{{Gair} et~al.}{2013}]{Gair2013}
{Gair} J.~R.,  {Vallisneri} M.,  {Larson} S.~L.,   {Baker} J.~G.,  2013,
  \mn@doi [Living Reviews in Relativity] {10.12942/lrr-2013-7}, \href
  {http://adsabs.harvard.edu/abs/2013LRR....16....7G} {16, 7}

\bibitem[\protect\citeauthoryear{{Graham} et~al.,}{{Graham}
  et~al.}{2015}]{Graham2015}
{Graham} M.~J.,  et~al., 2015, \mn@doi [\nat] {10.1038/nature14143}, \href
  {http://adsabs.harvard.edu/abs/2015Natur.518...74G} {518, 74}

\bibitem[\protect\citeauthoryear{{Haiman}, {Kocsis}  \& {Menou}}{{Haiman}
  et~al.}{2009}]{Haiman2009}
{Haiman} Z.,  {Kocsis} B.,   {Menou} K.,  2009, \mn@doi [\apj]
  {10.1088/0004-637X/700/2/1952}, \href
  {http://adsabs.harvard.edu/abs/2009ApJ...700.1952H} {700, 1952}

\bibitem[\protect\citeauthoryear{{Hiscock}, {Larson}, {Routzahn}  \&
  {Kulick}}{{Hiscock} et~al.}{2000}]{Hiscock2000}
{Hiscock} W.~A.,  {Larson} S.~L.,  {Routzahn} J.~R.,   {Kulick} B.,  2000,
  \mn@doi [\apjl] {10.1086/312867}, \href
  {http://adsabs.harvard.edu/abs/2000ApJ...540L...5H} {540, L5}

\bibitem[\protect\citeauthoryear{Hunter}{Hunter}{2007}]{Matplotlib}
Hunter J.~D.,  2007, \mn@doi [Computing In Science \& Engineering]
  {10.1109/MCSE.2007.55}, 9, 90

\bibitem[\protect\citeauthoryear{{Husa}, {Khan}, {Hannam}, {P{\"u}rrer},
  {Ohme}, {Forteza}  \& {Boh{\'e}}}{{Husa} et~al.}{2016}]{Husa2016}
{Husa} S.,  {Khan} S.,  {Hannam} M.,  {P{\"u}rrer} M.,  {Ohme} F.,  {Forteza}
  X.~J.,   {Boh{\'e}} A.,  2016, \mn@doi [\prd] {10.1103/PhysRevD.93.044006},
  \href {http://adsabs.harvard.edu/abs/2016PhRvD..93d4006H} {93, 044006}

\bibitem[\protect\citeauthoryear{Jones, Oliphant, Peterson  et~al.}{Jones
  et~al.}{2001}]{scipy}
Jones E.,  Oliphant T.,  Peterson P.,   et~al., 2001, {SciPy}: Open source
  scientific tools for {Python}, \url {http://www.scipy.org/}

\bibitem[\protect\citeauthoryear{Katz}{Katz}{2018}]{BOWIE}
Katz M.,  2018, Binary Observability With Illustrative Exploration (BOWIE),
  \url {https://github.com/mikekatz04/BOWIE}

\bibitem[\protect\citeauthoryear{{Kelley}, {Blecha}  \& {Hernquist}}{{Kelley}
  et~al.}{2017}]{PTA-illustris}
{Kelley} L.~Z.,  {Blecha} L.,   {Hernquist} L.,  2017, \mn@doi [\mnras]
  {10.1093/mnras/stw2452}, \href
  {http://adsabs.harvard.edu/abs/2017MNRAS.464.3131K} {464, 3131}

\bibitem[\protect\citeauthoryear{{Khan}, {Husa}, {Hannam}, {Ohme},
  {P{\"u}rrer}, {Forteza}  \& {Boh{\'e}}}{{Khan} et~al.}{2016}]{Khan2016}
{Khan} S.,  {Husa} S.,  {Hannam} M.,  {Ohme} F.,  {P{\"u}rrer} M.,  {Forteza}
  X.~J.,   {Boh{\'e}} A.,  2016, \mn@doi [\prd] {10.1103/PhysRevD.93.044007},
  \href {http://adsabs.harvard.edu/abs/2016PhRvD..93d4007K} {93, 044007}

\bibitem[\protect\citeauthoryear{{Klein} et~al.,}{{Klein}
  et~al.}{2016}]{Klein2016}
{Klein} A.,  et~al., 2016, \mn@doi [\prd] {10.1103/PhysRevD.93.024003}, \href
  {http://adsabs.harvard.edu/abs/2016PhRvD..93b4003K} {93, 024003}

\bibitem[\protect\citeauthoryear{{Kormendy} \& {Ho}}{{Kormendy} \&
  {Ho}}{2013}]{Kormendy2013}
{Kormendy} J.,  {Ho} L.~C.,  2013, \mn@doi [\araa]
  {10.1146/annurev-astro-082708-101811}, \href
  {http://adsabs.harvard.edu/abs/2013ARA%26A..51..511K} {51, 511}

\bibitem[\protect\citeauthoryear{{Kormendy} \& {Richstone}}{{Kormendy} \&
  {Richstone}}{1995}]{Kormendy1995}
{Kormendy} J.,  {Richstone} D.,  1995, \mn@doi [\araa]
  {10.1146/annurev.aa.33.090195.003053}, \href
  {http://adsabs.harvard.edu/abs/1995ARA%26A..33..581K} {33, 581}

\bibitem[\protect\citeauthoryear{{Larson}, {Hiscock}  \& {Hellings}}{{Larson}
  et~al.}{2000}]{Larson2000}
{Larson} S.~L.,  {Hiscock} W.~A.,   {Hellings} R.~W.,  2000, \mn@doi [\prd]
  {10.1103/PhysRevD.62.062001}, \href
  {http://adsabs.harvard.edu/abs/2000PhRvD..62f2001L} {62, 062001}

\bibitem[\protect\citeauthoryear{{Lin} et~al.,}{{Lin} et~al.}{2018}]{Lin2018}
{Lin} D.,  et~al., 2018, \mn@doi [Nature Astronomy]
  {10.1038/s41550-018-0493-1}, \href
  {http://adsabs.harvard.edu/abs/2018NatAs...2..656L} {2, 656}

\bibitem[\protect\citeauthoryear{{McConnell}, {Ma}, {Gebhardt}, {Wright},
  {Murphy}, {Lauer}, {Graham}  \& {Richstone}}{{McConnell}
  et~al.}{2011}]{McConnell2011}
{McConnell} N.~J.,  {Ma} C.-P.,  {Gebhardt} K.,  {Wright} S.~A.,  {Murphy}
  J.~D.,  {Lauer} T.~R.,  {Graham} J.~R.,   {Richstone} D.~O.,  2011, \mn@doi
  [\nat] {10.1038/nature10636}, \href
  {http://adsabs.harvard.edu/abs/2011Natur.480..215M} {480, 215}

\bibitem[\protect\citeauthoryear{{Miller}}{{Miller}}{2007}]{Miller2007}
{Miller} J.~M.,  2007, \mn@doi [\araa]
  {10.1146/annurev.astro.45.051806.110555}, \href
  {http://adsabs.harvard.edu/abs/2007ARA%26A..45..441M} {45, 441}

\bibitem[\protect\citeauthoryear{Moore, Cole  \& Berry}{Moore
  et~al.}{2015}]{Moore2015}
Moore C.~J.,  Cole R.~H.,   Berry C. P.~L.,  2015, Classical and Quantum
  Gravity, 32, 015014

\bibitem[\protect\citeauthoryear{{Petiteau}, {Babak}  \& {Sesana}}{{Petiteau}
  et~al.}{2011}]{Petiteau2011}
{Petiteau} A.,  {Babak} S.,   {Sesana} A.,  2011, \mn@doi [\apj]
  {10.1088/0004-637X/732/2/82}, \href
  {http://adsabs.harvard.edu/abs/2011ApJ...732...82P} {732, 82}

\bibitem[\protect\citeauthoryear{{Planck Collaboration} et~al.,}{{Planck
  Collaboration} et~al.}{2016}]{Planck2015}
{Planck Collaboration} et~al., 2016, \mn@doi [\aap]
  {10.1051/0004-6361/201525830}, \href
  {http://adsabs.harvard.edu/abs/2016A%26A...594A..13P} {594, A13}

\bibitem[\protect\citeauthoryear{{Rasskazov} \& {Merritt}}{{Rasskazov} \&
  {Merritt}}{2017a}]{Rasskazov1}
{Rasskazov} A.,  {Merritt} D.,  2017a, \mn@doi [\prd]
  {10.1103/PhysRevD.95.084032}, \href
  {http://adsabs.harvard.edu/abs/2017PhRvD..95h4032R} {95, 084032}

\bibitem[\protect\citeauthoryear{{Rasskazov} \& {Merritt}}{{Rasskazov} \&
  {Merritt}}{2017b}]{Rasskazov2}
{Rasskazov} A.,  {Merritt} D.,  2017b, \mn@doi [\apj]
  {10.3847/1538-4357/aa6188}, \href
  {http://adsabs.harvard.edu/abs/2017ApJ...837..135R} {837, 135}

\bibitem[\protect\citeauthoryear{{Reines}, {Greene}  \& {Geha}}{{Reines}
  et~al.}{2013}]{Reines2013}
{Reines} A.~E.,  {Greene} J.~E.,   {Geha} M.,  2013, \mn@doi [\apj]
  {10.1088/0004-637X/775/2/116}, \href
  {http://adsabs.harvard.edu/abs/2013ApJ...775..116R} {775, 116}

\bibitem[\protect\citeauthoryear{{Reis}, {Fabian}, {Ross}, {Miniutti}, {Miller}
   \& {Reynolds}}{{Reis} et~al.}{2008}]{Reis2008}
{Reis} R.~C.,  {Fabian} A.~C.,  {Ross} R.~R.,  {Miniutti} G.,  {Miller} J.~M.,
   {Reynolds} C.,  2008, \mn@doi [\mnras] {10.1111/j.1365-2966.2008.13358.x},
  \href {http://adsabs.harvard.edu/abs/2008MNRAS.387.1489R} {387, 1489}

\bibitem[\protect\citeauthoryear{{Robson} \& {Cornish}}{{Robson} \&
  {Cornish}}{2017}]{Robson2017}
{Robson} T.,  {Cornish} N.,  2017, preprint, \href
  {http://adsabs.harvard.edu/abs/2017arXiv170509421R} {} (\mn@eprint {arXiv}
  {1705.09421})

\bibitem[\protect\citeauthoryear{{Salcido}, {Bower}, {Theuns}, {McAlpine},
  {Schaller}, {Crain}, {Schaye}  \& {Regan}}{{Salcido}
  et~al.}{2016}]{Salcido2016}
{Salcido} J.,  {Bower} R.~G.,  {Theuns} T.,  {McAlpine} S.,  {Schaller} M.,
  {Crain} R.~A.,  {Schaye} J.,   {Regan} J.,  2016, \mn@doi [\mnras]
  {10.1093/mnras/stw2048}, \href
  {http://adsabs.harvard.edu/abs/2016MNRAS.463..870S} {463, 870}

\bibitem[\protect\citeauthoryear{{Schutz}}{{Schutz}}{1986}]{Schutz1986}
{Schutz} B.~F.,  1986, \mn@doi [\nat] {10.1038/323310a0}, \href
  {http://adsabs.harvard.edu/abs/1986Natur.323..310S} {323, 310}

\bibitem[\protect\citeauthoryear{{Sesana}, {Gair}, {Berti}  \&
  {Volonteri}}{{Sesana} et~al.}{2011}]{Sesana2011}
{Sesana} A.,  {Gair} J.,  {Berti} E.,   {Volonteri} M.,  2011, \mn@doi [\prd]
  {10.1103/PhysRevD.83.044036}, \href
  {http://adsabs.harvard.edu/abs/2011PhRvD..83d4036S} {83, 044036}

\bibitem[\protect\citeauthoryear{{V{\'e}ron-Cetty} \&
  {V{\'e}ron}}{{V{\'e}ron-Cetty} \& {V{\'e}ron}}{2010}]{Veron-Cetty2010}
{V{\'e}ron-Cetty} M.-P.,  {V{\'e}ron} P.,  2010, \mn@doi [\aap]
  {10.1051/0004-6361/201014188}, \href
  {http://adsabs.harvard.edu/abs/2010A%26A...518A..10V} {518, A10}

\bibitem[\protect\citeauthoryear{Walt, Colbert  \& Varoquaux}{Walt
  et~al.}{2011}]{Numpy}
Walt S. v.~d.,  Colbert S.~C.,   Varoquaux G.,  2011, \mn@doi [Computing in
  Science and Engg.] {10.1109/MCSE.2011.37}, 13, 22

\makeatother
\end{thebibliography}


\bsp	
\label{lastpage}
\end{document}